\documentclass[a4paper,10pt]{article}
\usepackage{amsmath}
\usepackage{graphicx}
\usepackage{epsfig}
\usepackage{here}
\newcommand{\beq}{\begin{equation}}
\newcommand{\eeq}{\end{equation}}
\newcommand{\beqa}{\begin{eqnarray}}
\newcommand{\eqa}{\end{eqnarray}}

\title{Time dependence of the survival probability of an opinion in a closed community}

\author{\centerline{Ekrem Ayd\i ner and Meltem Gonulol}\\
\textit{Department of Physics, Faculty of Arts and Sciences}\\
\textit{University of Dokuz Eyl\"{u}l, Izmir, Turkey} \\
\textit{E-mail: ekrem.aydiner@deu.edu.tr}}

\begin{document}

\maketitle

\section*{Abstract}
The time dependence of the survival probability of an opinion in a
closed community has been investigated in accordance with social
temperature by using the Kawasaki-exchange dynamics based on
previous study in Ref. [1]. It is shown that the survival
probability of opinion decays with stretched exponential law
consistent with previous static model. However, the crossover
regime in the decay of the survival probability has been observed
in this dynamic model unlike previous model. The decay
characteristics of both two regimes obey to stretched exponential.

\vspace*{25pt} \noindent {\bf Keywords:} Ising Model; Politics;
Random Walk; Sociophysics; Sznajd Model.

\newpage

\section{Introduction}
Binary models like Ising-type simulation have a long history. They
have been applied by Schelling to describe the ghetto formation in
the inner cities of the USA, i.e., to study phase separation
between black and white \cite{Schelling}. In the sociophysics
context, recently, many social phenomena such as election,
propagation of information, predicting features of traffic,
migration, opinion dynamics and formation in a social group have
been successful modelled based on Ising spin systems using models
and tools of statistical physics. With this respect, particularly
successful models have been developed by Sznajd \cite{Sznajd1},
Deffuant et al.\cite{Deffuant} and Hegselmann and Krause
\cite{Hegselmann}.

Among those three models, the one developed by Sznajd is the most
appropriate for simulation in networks and lattices, since it
consider just the interactions between the nearest neighbors.
Indeed, the Sznajd model has been successfully applied to model
sociophysical and economic systems \cite{Sznajd2}. On the other
hand, several modifications of the Sznajd model have been studied
using different rules or topologies starting from different
initial opinion densities \cite{Sznajd2,Fortunato,Stauffer1}. All
these models are static (i.e. not dynamic) and they allow for
consensus (one final opinion), polarization (two final opinion),
and fragmentation (more than two final opinions), depending on how
tolerant people are to different opinions.

More recently the striking sociophysical model has been suggested
by Ayd\i ner \cite{Aydiner} in order to explain the time evolution
of resistance probability of a closed community in a
one-dimensional Sznajd like model based on Ising spin system. It
has been shown that resistance probability in this model decay as
a stretched exponential with time. In that model spins does not
move on the lattice sites during the simulation, so this model was
so-called static. However, in a realistic case, spins i.e., people
move in the community i.e., in the space. Social or opinion
formation formed depend upon dynamics of the system. Because,
there must be a direct connection between opinion dynamics and
formation in a social system since the social formation is
determined by the dynamics. Meyer-Ortmanns  \cite{Meyer} studied
recent work in which the condition for ghetto formation in a
population with natives and immigrants by using Kawasaki-exchange
dynamics in a two dimensional Ising model. She showed that ghetto
formation can be avoided with a temperature increasing with time.
Similarly, Schulze have also generalized Meyer-Ortmanns work to up
to seven different ethnic groups to explain ghetto formation in a
multi-cultural societies in a Potts-like model \cite{Schulze}.

In this study, we have developed a dynamic version of the Ayd\i
ner \cite{Aydiner} model by combining the Ayd\i ner and
Meyer-Ortmanns \cite{Meyer} models based on one-dimensional Ising
model.

\section{Kinetic Model and Simulation}

In one-dimensional static model \cite{Aydiner}, each site carriers
a spin which is either spin up (+1) or spin down (-1) randomly.
Spin up (+1) represent the host people and spin down (-1)
represent the soldier. The host people always against occupation,
and, on the other hand, soldier always willing to continue
occupation, who always have the opinion opposite of that of the
host people. Furthermore, the community member i.e., spins doesn't
also move on the lattice during the process.

In this model, initially, it was assumed that there was a over all
consensus among member of the community against occupation even if
some exceptions exist. One expects that host people obey to this
consensus at least initially. In this sense, community behaves as
polarized at zero social temperature \cite{Schweitzer} against
occupation just like Ising ferromagnet at zero temperature.

It was conjectured that host people are influenced by soldiers
even though they against occupation owing to they are exposed to
intensive biased information or propagation. Soldiers affect the
host people and force to change their opinion about occupation.
Effected people may change their own opinions depending on
resistance probability of the nearest neighbors about occupation.
Moreover, effected host people affect neighbors. Such a mechanism
depolarize the polarization (resistance probability) of all host
people. Hence social polarization destroy.

However, soldiers, unlike host people, have not been influenced by
the host people. Their opinion about justifying the occupation
does not change during the occupation process, since they may be
stubborn, stable or professional etc., who behaves like persistent
spins in Ising spin system. It is means that the probability of
the against occupation of a soldier is always zero.

If we summarize, we can say that none spins does flip fully in the
system. Spin up always remains spin up, and spin down always
remains spin down. In this respect, the probability of against
occupation of host people can be interpreted as a survival
probability of opinion of host people about occupation under above
considerations. In this sense, the survival probability $W_{i}$ of
opinion of host people indicate equal to $1$ at least initially
and, on the other hand, the probability of against occupation of
soldier equal to zero, which means that soldier behaves as a trap
point lattice which depolarize the survival probability of opinion
of host people.

Of course, one may suggest that there are many different number of
opinions in society, however, it is possible to find that a
society being formed two-state opinion in a real case. Therefore
this model is a good example for two-state opinion model as well
Galam contrarian model \cite{Galam} even though it seems that it
is very simple. Furthermore, in real social systems, people move
on the space, i.e., lattice. Therefore, in this study, we assumed
that people i.e., spins randomly move on the lattice through the
Kawasaki-exchange dynamics contrary to previous model.

The survival probability $W_i$ for a people at site $i$ at the
next time $t+1$ is determined with the survival probability of
nearest-neighbors with previous time $t$ as
\begin{equation}
W_{i}(t+1)=\frac{1}{2}[W_{i+1}(t)+W_{i-1}(t)].\label{eq1}
\end{equation}
We note that the survival probability for all site are calculated
as synchronously.

Randomly motion of the spins i.e., people on the lattice through
the Kawasaki-exchange dynamics. Firstly, a spin pair is chosen
randomly and then it is decided whether spin pair exchange with
each other or not. In this approach, the nearest-neighbor spins
are exchanged under heat-bath dynamics, i.e., with probability
$p\sim\exp\left(-\Delta E/k_{B}T\right))$, where $\Delta E$ is the
energy change under the spin exchange, $k_{B}$ is the Boltzmann
constant, and $T$ is the temperature i.e., social temperature or
tolerance. Hence, to obtain probability $p$ we need to calculate
$E_1$ and $E_2$ which correspond to energy of the spin pair at
first position and after exchange with position of spins,
respectively. Energy $E_1$ and $E_2$ can be calculated in terms of
the survival probability instead of spin value as
\begin{subequations}
\begin{equation}
E_{1}(t)=aW_{i}(t)+bW_{i+1}(t) \label{eq2a}
\end{equation}
\begin{equation}
E_{2}(t)=aW_{i+1}(t)+bW_{i}(t) \label{eq2b}
\end{equation}
\end{subequations}
where
$$a=[W_{i-1}(t)+W_{i+1}(t)]$$
and
$$b=[W_{i}(t)+W_{i+2}(t)]$$
Energy difference is written as $\Delta E=E_2-E_{1}$ from
Eq.~(\ref{eq2a}) and (\ref{eq2b}).

In addition, the total survival probability of opinion of host
people at the any time $t$ can be obtained over each person for
any $r$ configuration as
\begin{equation}
P_{r}(t) =\frac{1}{m_{0}} {\sum_{i}}  W_{i}(t) \label{eq3}
\end{equation}
where $m_{0}$ is the initial number of host people. On the other
hand, the averaged survival probability at the any time $t$ can be
obtained from Eq.~(\ref{eq3}) over the independent configuration
as
\begin{equation} \label{eq4}
<P(t) > =\frac{1}{R} {\sum_{r=1}^{R}}P_{r}(t)
\end{equation}
where $R$ is the number of different configurations.
\begin{figure}
\centerline{\psfig{file=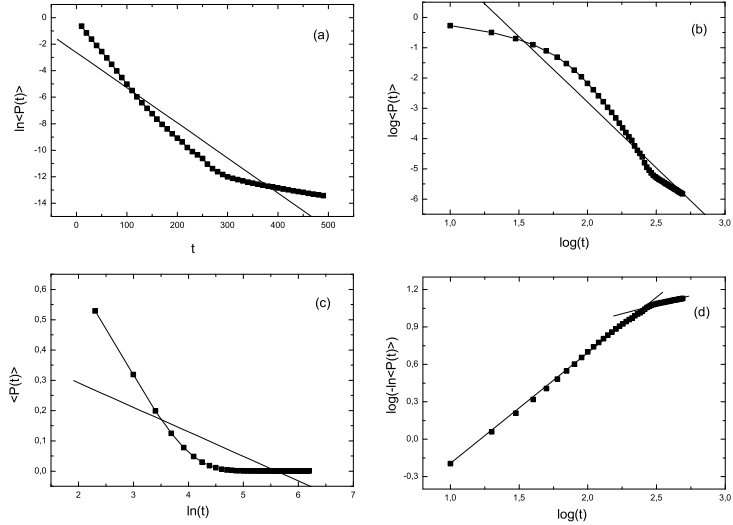}} \caption{Data for
$\rho=0.1$ were plotted $\ln<P(t)>$ versus $t$ in
Fig.~\ref{fig1}(a), $\log<P(t)>$ versus $\log t$ in
Fig.~\ref{fig1}(b), $<P(t)>$ versus $\ln t$ in Fig.~\ref{fig1}(c),
and $\log(-\ln<P(t)>)$ versus $\log t$ in Fig.~\ref{fig1}(d).
Solid-dot lines indicate data and solid lines represent fitting
curves in all figures. \label{fig1}}
\end{figure}

\section{Results and Discussion}

We have adopted the Monte Carlo simulation technique to the
one-dimensional sociophysical model using the lattice size
$L=1000$ with periodic boundary condition, and independent
configuration $R=1000$ for the averaged results. The simple
algorithm for the simulation is as follows: i) at the $t=0$,
Eq.~(\ref{eq4}) is initially calculated, ii) for $t>0$ a spin pair
is randomly chosen, and then it is decided whether the spin pair
exchange or not with the probability $p\sim\exp\left(-\Delta
E/k_{B}T\right))$, this step is repeated $L$ times, iii) after
ii-steps are completed, Eq.~(\ref{eq4}) is recalculated again, and
to continue this procedure goes to step ii.
\begin{figure}
\centerline{\psfig{file=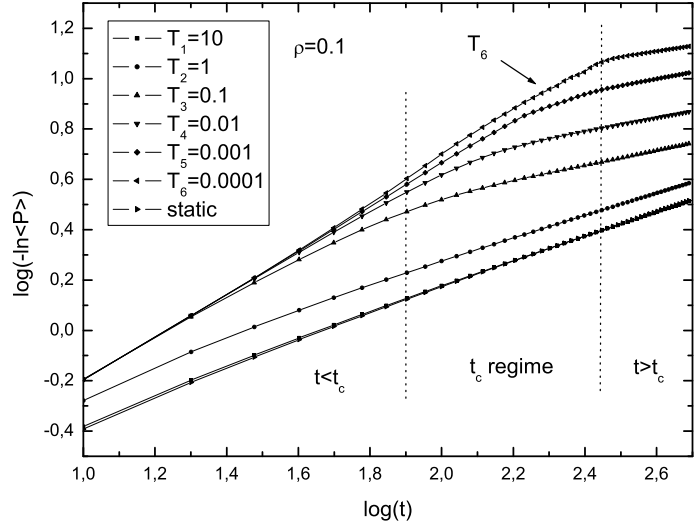}} \caption{The time
dependence of the survival probability of the opinion of host
people decays KWW i.e., stretched exponential with time for
different social temperature $T$. The time crossover appears in
the time evolution of survival probability of the opinion at low
social temperatures. The crossover becomes more clear when social
temperature decreases. \label{fig2}}
\end{figure}
\begin{figure}
\centerline{\psfig{file=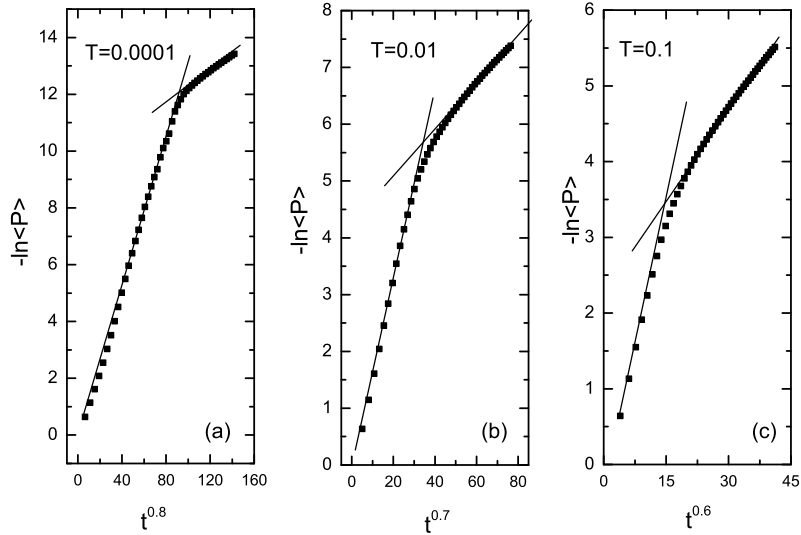}} \caption{(a) $-\ln<P(t)>$
versus $t^{0.8}$ for $T=0.0001$, (b) $-\ln<P(t)>$ versus $t^{0.7}$
for $T=0.01$, and (c) $-\ln<P(t)>$ versus $t^{0.6}$ for $T=0.1$.
All figures are plotted for fixed $\rho$ value i.e. $\rho=0.1$,
and solid lines represent fitting curves. \label{fig3}}
\end{figure}
\begin{figure}
\centerline{\psfig{file=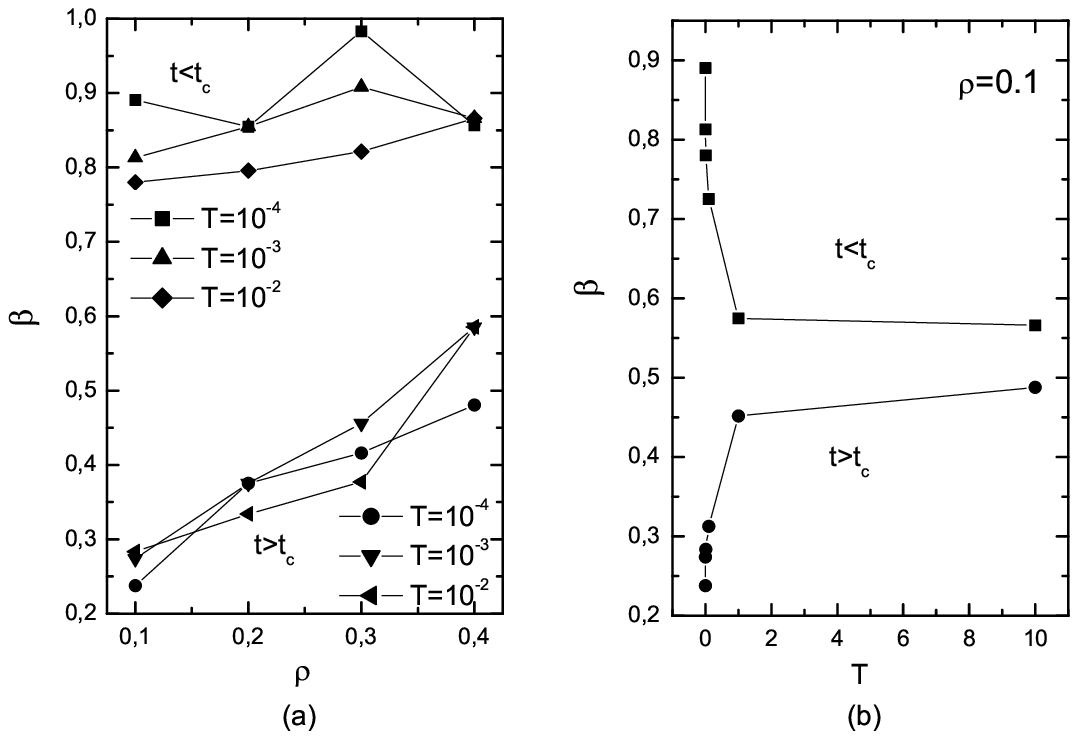}} \caption{(a) The soldier
density dependence of the exponent $\beta$ under and below the
crossover time $t_c$ for different social temperatures, (b) Change
of the decay exponent $\beta$ depend on social temperature $T$
under and below of the crossover time $t_c$ for a fixed soldier
density $\rho=0.1$. \label{fig4}}
\end{figure}

The simulation results are as follow: We have firstly plotted
simulation data versus time in Fig.~\ref{fig1} in a several
manner. It is explicitly seen from Figs.~\ref{fig1}(a)-(c) that
there are no power, exponential and logarithmic law dependence in
our simulation data, respectively. However, as seen
Fig.~\ref{fig1}(d), data well fit to the stretched exponential
function as
\begin{equation} \label{eq5}
<P(t)> \sim e^{-\lambda t^{\beta}}%
\end{equation}
where $\lambda$ is the relaxation constant, and $\beta$ is the
decay exponent of the survival probability.  This result indicate
that the time evaluation of survival probability of the opinion of
the host people in a closed community has stretched exponential
character i.e., Kohlraush-William-Watts (KWW) decay law
\cite{Kohlraush,WilliamWatts}.

It should be tested whether Fig.~\ref{fig1}(d) satisfies to
stretched exponential or not \cite{Stauffer2}. Because, as noted
by Stauffer, the Fig.~\ref{fig1}(d) would work as stretched
exponential, if pre-factor of Eq.~(\ref{eq5}) is equal to 1.
However, if pre-factor is less than $1$, it may give the
impression of stretched exponential form, even for $\beta=1$.
Therefore, it can be plotted $-\ln<P(t)>$ versus suitable powers
of $t$, like $t^{1/2}$, $t^{1/3}$, etc., and find out the best
straight line among the powers of $t$ for long times. Hence,
$-\ln<P(t)>$ was plotted versus powers of $t$ for $\rho=0.1$ then
the best straight fitting line for long times was obtained for
$\beta=0.8$ for $T=0.0001$, $\beta=0.7$ for $T=0.01$, and
$\beta=0.6$ for $T=0.1$ as seen in Fig.~\ref{fig3}(a)-(c)
respectively. These results confirm to this method used to find
out stretched exponential exponents in Fig.~\ref{fig1}(d), and
also all figures in Fig.~\ref{fig2} as mentioned . Also, this test
indicates that prefactor in Eq.~(\ref{eq5}) does not effect
results presented in this paper.
\begin{figure}
\centerline{\psfig{file=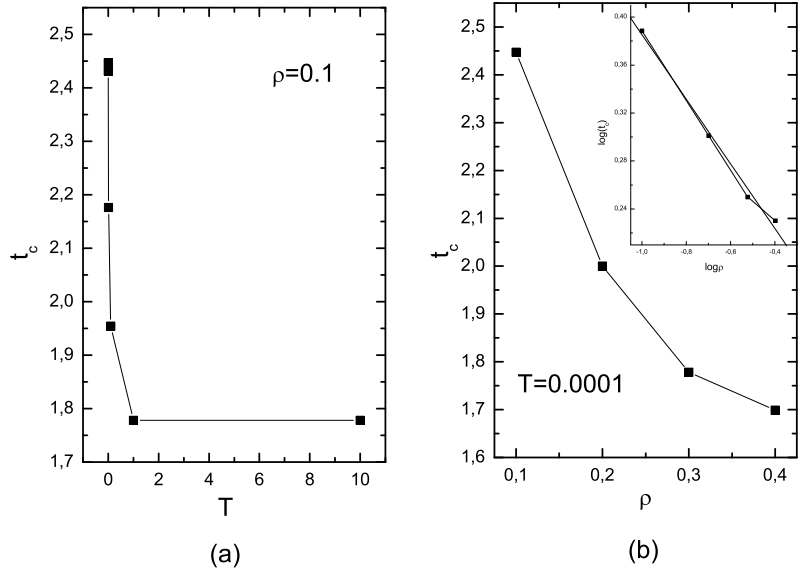}} \caption{(a) The social
temperature dependence of the crossover time $t_c$ for fixed
soldier density $\rho=0.1$, (b) The soldier density dependence of
the crossover time $t_c$ for fixed social temperature $T=0.0001$.
\label{fig5}}
\end{figure}

It is concluded that results for high temperatures also consistent
with static model \cite{Aydiner}. But, unlike the static model,
time crossover has been observed in dynamic model at low
temperatures. In order to investigate the transition we have
plotted survival probability versus time for different social
temperature $T$ in Fig.~\ref{fig2}. It is clearly seen that the
time crossover occurs depend on social temperature. When social
temperature decreases, the crossover become more clear. Such a
behavior was not observed in a static model. We can bridge the
short time regime and the long time regime by a scaling function
$f\left(t/t_{c}\right)$
\begin{equation} \label{eq6}
\left\langle P\left(  t\right)  \right\rangle = e^{-\left(
t/\tau\right) ^{\beta}}f\left(  t/t_{c}\right)
\end{equation}
where $t_{c}$ indicates the time crossover. For our simulation
data, the scaling relation (6) can be written for very long and
very short time intervals as
\begin{equation} \label{eq7}
\left\langle P\left(  t\right)  \right\rangle \sim \Big{\{}
\begin{array}
[c]{c}%
e^{-\left(  t/\tau\right)  ^{\beta_{1}}}\\
e^{-\left(  t/\tau\right)  ^{\beta_{2}}}%
\end{array}
\begin{array}
[c]{l}%
if \hspace{0.3cm} t<<t_{c}\\
if \hspace{0.3cm} t>>t_{c}.%
\end{array}
\end{equation}

On the other hand, in order see how the decay exponent $\beta$
depend on soldier density $\rho$, and social temperature $T$, we
have plotted $\beta$ versus soldier density $\rho$ in
Fig.~\ref{fig4}(a) for $t<t_c$ and $t>t_c$ in account to taken
different social temperatures, and social temperature $T$ in
Fig.~\ref{fig4}(b) for a fixed value of density $\rho$,
respectively.

As seen from Fig.~\ref{fig4}(a) that $\beta_1$ and $\beta_2$ are
linearly depend on soldier density both of two regimes at low
social temperature. On the other hand, the decay exponent has two
different character for $t<t_c$ and $t>t_c$ depend on social
temperature $T$ in Fig.~\ref{fig4}(b), the decay exponent
$\beta_1$ decreases with increasing temperature $T$ for $t<t_c$,
whereas $\beta_2$ increases with increasing temperature $T$ for
$t>t_c$ at low temperatures. However, for relatively high
temperatures we roughly say that $\beta_1$ approach to $\beta_2$
for both two regimes obey to Eq.~(\ref{eq7}).

Finally, to understand the social temperature and soldier density
dependence of the time crossover $t_c$, we have plotted $t_c$
versus social temperature $T$ in Fig.~\ref{fig5}(a) for a fixed
soldier density $\rho$, and versus soldier density in
Fig.~\ref{fig5}(b) for fixed social temperature $T$, respectively.
It seems from Fig.~\ref{fig5}(a) that the crossover transition
$t_c$ quite rapidly decrease with increasing $T$, on the other
hand, it seems from Fig.~\ref{fig5}(b) that it slowly decrease
with increasing soldier density $\rho$. We note that as seen
inserted figure in Fig.~\ref{fig5}(b) the crossover transition
$t_c$ depends on soldier density with power law for fixed social
temperature.

\section{Conclusions}

We suggest that the stretched exponential behavior of decay must
be originated from model system. The persistent spins i.e., the
soldiers doesn't flip during simulation, therefore they behave as
a trap in the system. Hence they play a role diminishing the
survival probability of the neighbor spins in the system.
Consequently, decay characteristic of the system can be explain
due to the trapping states. Another say, this characteristic
behavior doesn't depend on either diffusion dynamics of spins or
interaction rules between spins.

Another unexpected behavior is the time crossover in $\beta$
contrast to previous model \cite{Aydiner}. We supposed that this
amazing result originated from opinion dynamics depend on social
temperature. Model allows to the opinion formation with time.
Indeed, there is a direct connection between opinion dynamics and
formation in a social system since the social formation is
determined by the dynamics as depend on the social temperature.
For example, in a real spin system, decreasing temperature phase
separation may occur in the system. In the sociophysical sense, it
means that people who have different opinion are separated each
other with decreasing social tolerance, and therefore the ghetto
formation or polarization may occur in the system.

It is expected that interactions between soldier and host people
is maximum when soldiers are randomly distributed in the
community. As social temperature, i.e., tolerance is decreased,
however, phase separation occur with time, so this leads to
decreasing of the interactions.

In our opinion, the ghetto formation in the system doesn't leads
crossover transition in time because of the ghetto formation is
randomly distributed relatively. On the other hand, the time
average of survival probability over different configuration
effect of ghetto formation may probably destroy. So we don't hope
that ghetto formation is not responsible crossover transition.
However, polarization must be occurred at low temperature leads to
meaningful phase separation in the system. Such a polarization may
leads to crossover transition in time.

Stretched exponential behavior indicates mathematically that decay
for the relatively short times is fast, but for relatively long
times it is slower. One can observe that this mathematical
behavior corresponds to occupation processes in the real world. In
generally, a military occupation is realized after a hot war. The
community does not react to occupation since it occurs as a result
of defeat. People are affected easily by propaganda or other
similar ways. Therefore, it is not surprised that resistance
probability decrease rapidly at relatively short times. On the
other hand, spontaneous reaction may begin against occupation in
the community after the shock. Hence, community begins by
regaining consciousness and more organized resistance may display
difficulties for occupants. For long times, the resistance
probability decreases more slowly. This means that resistance
against occupation extends to long times in practice. At this
point, the number of soldiers is also important, because the
density of soldiers determines the speed of decaying.

The different regimes have been observed in the decay of the
survival probability. These regimes clearly appear particularly at
low temperatures. In the case of the social temperature is very
low, $\beta_1$ is bigger than $\beta_2$ which indicates the decay
of the survival probability for relatively short time is slower
than for relatively long time. This can be interpreted that the
resistance of host people against occupation may be broken
spontaneously if soldier can wait enough time.

Of course, the mechanism considered in this work can be regarded
as simple, but, it would be useful to understand the time
evolution of the resistance probability of the community against
to occupation in the one-dimensional model under some
considerations. We remember that simple social rules lead to
complicated social results, hence we believe that the obtained
results and model can be applied the real social phenomena in the
societies to understand the basis of them.

\section*{Acknowledgments}
Authors are grateful Dietrich Stauffer for the suggestions in the
preparation of this paper.

\end{document}